\documentclass[preprint,preprintnumbers,fleqn,amsmath,amssymb,aps]{revtex4}
\usepackage{graphicx}% Include figure files
\usepackage{time}
\usepackage{dcolumn}% Align table columns on decimal point
\usepackage{bm}% bold math %TCIDATA{OutputFilter=LATEX.DLL}
\usepackage{amsmath}

\begin{document}
\title{Suppression of the superconducting critical current of $\rm{Nb}$  in bilayers of
$\rm{Nb/SrRuO_3}$}
\author{M. Feigenson}
\altaffiliation{E-mail: faigenm@mail.biu.ac.il Tel: 972-3-5317629
Fax: 972-3-5353298}
\author{L. Klein}
\affiliation{Department of Physics, Bar Ilan University, Ramat Gan
52900, Israel}
\author{M. Karpovski}
\affiliation{School of Physics and Astronomy, Tel Aviv University,
Tel Aviv 69978, Israel}
\author{J. W. Reiner}
\author{M. R. Beasley}
\affiliation{Department of Applied Physics, Stanford University,
Stanford, California 94305}

\date{\today}
\maketitle

%\begin {abstract}
\begin{center}
\large Abstract
\end{center}
\normalsize \vspace{2 pt}

 In bilayers consisting of ferromagnetic and
superconducting films, the ferromagnetic film in its domain state
induces inhomogeneous distribution of magnetic fields in the
superconducting film. When the ferromagnetic film has bubble
magnetic domains in a labyrinth structure, it has been found that
the pinning of the vortices increases; hence, the critical current
of the superconducting film becomes larger. Here we study the
effect of parallel ferromagnetic domain structure in
$\rm{Nb/SrRuO_3}$ on the critical current of Nb with current
flowing perpendicularly to the domains and find that in this case
the ferromagnetic domain structure decreases the critical current.
%\end{abstract}
\newpage

\section{Introduction:}

\maketitle The critical current ($\rm{I_c}$) in type II
superconductors is limited by the pinning strength of the
vortices, since the current applies force on the vortices leading
to vortex motion which dissipates energy and destroys
superconductivity. Different techniques for creating vortex
pinning were proposed, including thickness modulation
\cite{thickness}, columnar defects \cite{col_dif} and introduction
of magnetic particles on the superconducting film
\cite{magn_part1,magn_part2}. Most of those techniques were based
on local suppression of the superconducting state, since it is
favorable for the vortex normal core to be located in areas which
are normal anyway; hence saving the condensation energy.

Recently it was proposed to pin the magnetic flux of vortices
rather than their normal core \cite{Bulaevskii}. This magnetic
pinning can be achieved in superconductor - ferromagnet (SC/FM)
multilayers. The magnetostatic interaction between magnetic flux
of the vortices and magnetization of the FM layer gives rise to
pinning potential across the domain structures that increases the
superconducting critical current \cite{Jan, Garcia-Santiago,
Lange}. This method was demonstrated in systems where the FM film
had bubble magnetic domains in a labyrinth structures, such as
$\rm{CoPt}$ or $\rm{BaFe_{12}O_{19}}$ \cite{Jan, Garcia-Santiago,
Lange}. On the other hand, no additional pinning force is expected
in FM/SC multilayers if the vortex motion is parallel to the
domain structure \cite{Jan}.

Here we study the influence of in-plane magnetic stripe domain
structure of $\rm{SrRuO_3}$ on the critical current of Nb in SC/FM
bilayers with current flowing perpendicularly to the domain
structure (Fig. 1). In this configuration vortices are forced to
move $\textit{parallel}$ to the domain walls due to the action of
Lorentz force $\rm[\textbf{F}=q\textbf{v}\times \textbf{B}]$. We
show that in this configuration the superconducting critical
current decreases, suggesting lower vortex pinning.

\section{Experiment:}
$\rm{SrRuO_3}$ is a 4d itinerant ferromagnet. In this experiment
we use thin films of $\rm{SrRuO_3}$ with Curie temperature of
$\rm{\sim150}$ K and saturated magnetic moment of $\thicksim$ 1.4
$\mu_B$ per ruthenium. The films have orthorhombic structure and
since they are grown on slightly miscut substrate of (001)
$\rm{SrTiO_3}$, they grow with their c axis in the film plane and
the a and b axis at $45^{0}$ out of the plane of the film. The
films have large uniaxial magnetocrystalline anisotropy field
($\rm{\sim10}$ T) roughly along the b axis and consequently the
magnetic domain structure is in form of stripes parallel to the
in-plane projection of the b-axis (see Figure 1) \cite{Marshall}.
The thickness of the domain walls is $\thicksim$ 3 nm and the
domain wall spacing is $\thicksim$ 200 nm.

The large uniaxial anisotropy field combined with the much smaller
self field ($\rm{4\pi M\sim0.2}$ T) prevent the creation of
closure domains and once the domain structure is annihilated by
applying sufficiently large magnetic field (2 T) at a temperature
lower than the Curie temperature, no new domains renucleate when
the field is set back to zero \cite{Marshall}.

Our measurements were done on SC/FM bilayers consisting of an
epitaxial thin film of $\rm{SrRuO_3}$ (900 {\AA}) grown by
reactive electron beam coevaporation \cite{Reiner} with properties
as described above and a thin policrystalline film of Nb (600
{\AA}) grown by sputtering. The layers of $\rm{SrRuO_3}$ and Nb
were separated with a buffer layer of Cu (20 {\AA}) to avoid
oxygen migration.

\begin{figure} [h!]
\begin{center}
\includegraphics {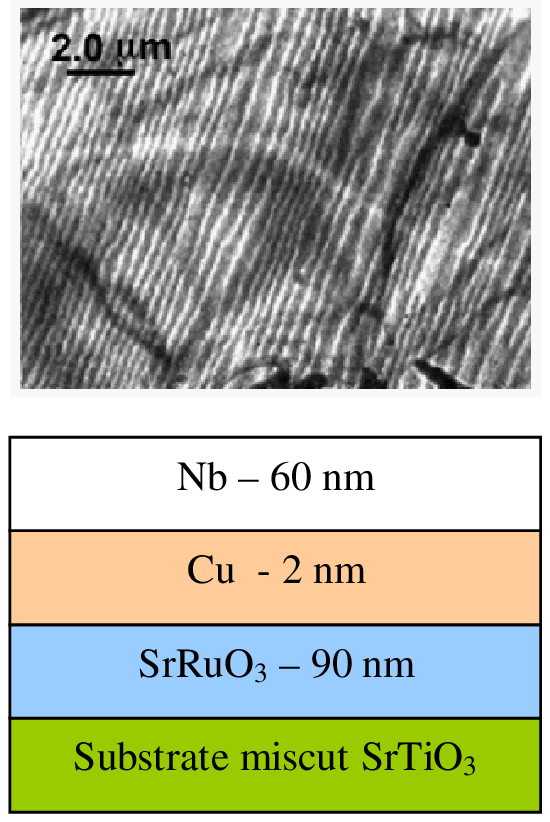}
\end{center}
\caption{Top: Image from \cite{Marshall} of domain walls in
$\rm{SrRuO_3}$ at 100 K with Lorentz mode TEM. Bright and dark
lines image domain walls at which the electron beam diverges or
converges, respectively. Bottom: FM/SC bilayer structure (with a
buffer layer of Cu).} \label{sample}
\end{figure}

To elucidate the effect of the domain structure in $\rm{SrRuO_3}$
on the critical current of Nb, I-V curves at zero applied field
were measured with the $\rm{SrRuO_3}$ film uniformly magnetized or
with the $\rm{SrRuO_3}$ film in its stripe domain structure. When
$\rm{SrRuO_3}$ is fully magnetized there is practically no induced
field on the Nb film due to demagnetization, so comparing the
critical current in the two cases yields the effect of the field
due to the domain structure.

To measure the I-V curves of the Nb film with no field penetrating
from the $\rm{SrRuO_3}$ film, the sample was cooled from above the
Curie temperature of $\rm{SrRuO_3}$ ($\rm{\thicksim 150}$ K) in a
field of 2 T down to 10 K. This prevented the formation of
magnetic domains in the $\rm{SrRuO_3}$ film. At 10 K the field was
set to zero and the $\rm{SrRuO_3}$ remained fully magnetized in a
single domain state. The sample was then cooled down under the
$\rm{T_c}$ of Nb (7 K) at zero field. The I-V curves were measured
at different temperatures from 2 K up to the $\rm{T_c}$ of Nb. We
call those measurements field cooled measurement (FC). Between
every two sequential I-V measurements the sample was heated above
the $\rm{T_c}$ of Nb so that each measurement started with no
penetration of vortices into the Nb film. In the second type of
measurements the sample was cooled from above the Curie
temperature of $\rm{SrRuO_3}$ at zero field to below the
$\rm{T_c}$ of Nb and I-V curves were measured at the same range of
temperatures. We call those measurements zero field cooled
measurement (ZFC). In those measurements magnetic domains formed
inside the $\rm{SrRuO_3}$ film and magnetic flux penetrated into
the Nb film.

\section{Results and Discussion:}

Figure 2 shows ZFC and FC I-V curves of $\rm{Nb/SrRuO_3}$ bilayers
measured at T=4.7 K. A strong suppression of $\rm{I_c}$ is
observed when $\rm{SrRuO_3}$ is in its domain state. The ZFC -
$\rm{I_c}$ obtained at 4.7 K is $\sim1.6$ times smaller than FC -
$\rm{I_c}$.

\begin{figure} [h!]
\begin{center}
\includegraphics [scale=0.5]{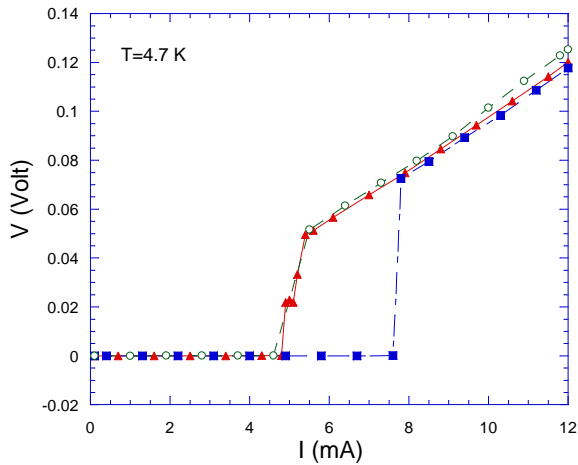}
\end{center}
\caption{Three I-V curves of a $\rm{Nb/SrRuO_3}$ bilayer at T=4.7
K: (a) $\rm{SrRuO_3}$ is in its domain structure (triangles); (b)
$\rm{SrRuO_3}$ is fully magnetized (squares); (c) $\rm{SrRuO_3}$
is fully magnetized and a magnetic field of H=1600 Oe is applied
along the plane of the sample (circles).} \label{I_V_T=4_7Kb}
\end{figure}

\begin{figure} [h!]
\begin{center}
\includegraphics [scale=0.5]{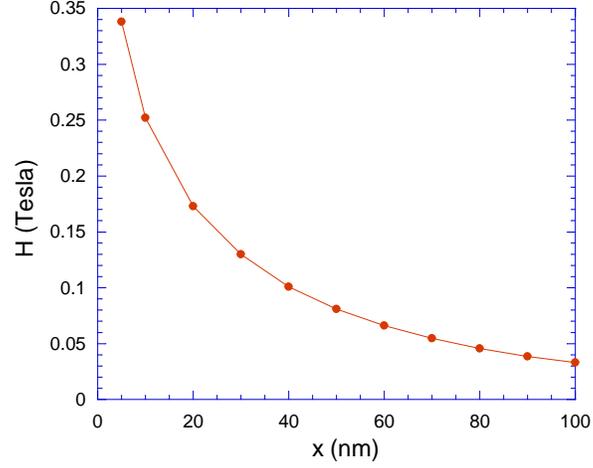}
\end{center}
\caption{Numerical calculations of the magnetic field induced by
the domain structure as a function of distance from the domain
wall, x, in our SC/FM bilayers, based on the calculations in Ref.
\cite{Sonin}. } \label{Hx}
\end{figure}

To estimate the value of the effective magnetic field induced by
the magnetic domain structure of $\rm{SrRuO_3}$ on the Nb film we
measured I-V curve with the $\rm{SrRuO_3}$ film fully magnetized
and external magnetic field, \textbf{H}, applied in the plane of
the sample. As it is seen in Figure 2, the effect of the domain
structure of $\rm{SrRuO_3}$ on the $\rm{I_c}$ of Nb is similar to
that of an applied magnetic field of 1600 Oe. Such a field is
consistent with our estimations based on calculations of Sonin
\cite{Sonin} (Fig 3), who calculated the fringing field created by
the magnetic structure at the interface between FM and SC layers
in the SC/FM bilayers. Despite the difference in the field
distribution in case of the field produced by the stripe domain
structure and the uniform applied field, we note that the results
are of the same order of magnitude.

\begin{figure} [h!]
\begin{center}
\includegraphics [scale=0.5]{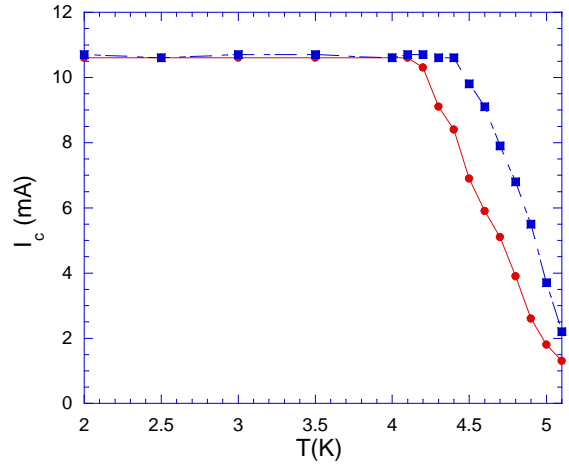}
\end{center}
\caption{$\rm{I_c}$ vs T of a $\rm{Nb/SrRuO_3}$ bilayer, when
$\rm{SrRuO_3}$ is fully magnetized (squares) and when it is in its
domain structure (circles).} \label{Ic_vs_T}
\end{figure}

Figure 4 displays the temperature dependence of the
superconducting critical current with the $\rm{SrRuO_3}$ film
fully magnetized and with the $\rm{SrRuO_3}$ film in stripe domain
structure. As it appears, there is a range of temperatures,
approximately between 4.2 and 5.2 K, where $\rm{I_c}$ is
suppressed by the magnetic field induced by the magnetic structure
of the $\rm{SrRuO_3}$ film. The suppression of $\rm{I_c}$ is less
clear above 5.2 K, since no sharp superconducting - normal phase
transition is observed and it is difficult to determine the
difference between the values of ZFC - $\rm{I_c}$ and FC -
$\rm{I_c}$.

\begin{figure} [h!]

\begin{center}
\includegraphics [scale=0.5]{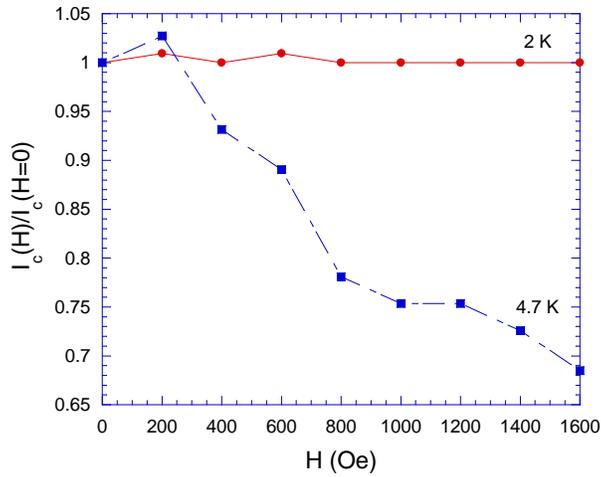}
\end{center}
\caption{$\rm{I_c}$ vs H of Nb at T=2 and 4.7 K normalized to the
values of $\rm{I_c}$ at H=0.} \label{Ic_H}
\end{figure}

In the temperature range with sharp superconductivity to normal
transition, the ZFC - $\rm{I_c}$ and FC - $\rm{I_c}$ are
practically identical at low temperatures, while above 4.2 K there
is a clear difference between their values. These behavior can be
understood from the field dependence of $\rm{I_c}$ at various
temperatures [see Fig. 5]. The I-V curves, measured at T=2 and 4.7
K, show the dependence of $\rm{I_c}$ on the magnetic field,
\textbf{H}, applied along the plane of the sample, when the
$\rm{SrRuO_3}$ film is fully magnetized. At low temperatures (T=2
K) no variations in $\rm{I_c}$ are observed in the field range
that domain walls can produce, while significant variations are
observed at 4.7 K for the same field range.

\section{Conclusion:}
We have shown that in bilayers of $\rm{Nb/SrRuO_3}$ with current
flowing perpendicularly to the domain walls there is a temperature
range in which clear suppression of the critical current is
observed.

\begin{acknowledgments}
L.K. acknowledges support by the Israel Science Foundation founded
by the Israel Academy of Sciences and Humanities.
\end{acknowledgments}

\end{document}